\documentclass[pra,twocolumn,showpacs]{revtex4-1}% Physical Review A

\usepackage{epsfig}
\usepackage{bm}% bold math
\usepackage[matrix,frame,arrow]{xy}

\usepackage[applemac]{inputenc}
\usepackage[T1]{fontenc}
\usepackage[english]{babel}
\usepackage{amsmath}
\usepackage{ae}
\usepackage{units}
\usepackage{amssymb}
\usepackage{icomma}
\usepackage{color}
\usepackage{graphicx}
\usepackage{bbm}
\usepackage{url}
\usepackage{nomencl}
\usepackage{subfigure}

\newcommand{\rd}{\ensuremath{\mathrm{d}}}
\newcommand{\id}{\ensuremath{\,\rd}}

\newcommand{\bra}[1]{\langle #1|} %\langle j \vbar
\newcommand{\ket}[1]{|#1\rangle}  %\vbar j \rangle \langle j \vbar

\newcommand{\brakket}[3]{\left\langle #1\left| #2 \right| #3\right\rangle}
\newcommand{\expec}[1]{\left\langle #1 \right\rangle}
\newcommand{\tr}[1]{\text{tr}\left( #1 \right)}
\newcommand{\expecc}[1]{\text{E}\left[#1\right]}

\newcommand{\comm}[2]{\left[ #1, #2 \right]}
\newcommand{\lind}[1]{\mathcal{D}\left[#1\right]}
\newcommand{\meas}[1]{\mathcal{M}\left[#1\right]}
\newcommand{\measb}[1]{\bar{\mathcal{M}}\left[#1\right]}
\newcommand{\measg}[1]{\mathcal{G}\left[#1\right]}
\newcommand{\measgb}[1]{\bar{\mathcal{G}}\left[#1\right]}

\newcommand{\abs}[1]{\left|#1\right|}

\newcommand{\rhoq}{\rho^\text{qb}}
\newcommand{\rhob}{\bar{\rho}}
\newcommand{\rhobq}{\bar{\rho}^\text{qb}}
\newcommand{\rhobp}{\bar{\rho}^\mathbf{P}}
\newcommand{\bP}{\mathbf{P}}

\newcommand{\Pia}{\Pi_\alpha}
\newcommand{\Piad}{\Pi_\alpha^\dag}
\newcommand{\Na}{N_\alpha}

\newcommand{\im}{\text{Im}}
\newcommand{\re}{\text{Re}}

\newcommand{\figref}[1]{\mbox{Fig.~\ref{#1}}}

\newcommand{\secref}[1]{\mbox{Sec.~\ref{#1}}}

\renewcommand{\eqref}[1]{\mbox{Eq.~(\ref{#1})}}

\newcommand{\be}{\begin{equation}}
\newcommand{\ee}{\end{equation}}
\newcommand{\dagg}[1]{#1^\dagger}

\usepackage[colorlinks]{hyperref}
\hypersetup{%
        plainpages=true,
        breaklinks=true,% not default in dvips mode, so we must specify
        hypertexnames=false,%not ideal, but needed when pagenums duplicate (`i' vs. `1')
        pageanchor=true,
        colorlinks=true,
        linkcolor={blue},
        citecolor={green},
        urlcolor={black},
        pagecolor={black},
        anchorcolor={black}
      }

\begin{document}

\title{Undoing measurement-induced dephasing in circuit QED}

\author{A. \surname{Frisk Kockum}}
%\email[e-mail:]{friska@chalmers.se}
\author{L. Tornberg}
\author{G. Johansson}

\affiliation{Department of Michrotechnology and Nanoscience$,$ MC2$,$\\ Chalmers\:University\:of\:Technology$,$ SE-412 96\:Gothenburg$,$ Sweden}

\date{\today}

\begin{abstract}
We analyze the backaction of homodyne detection and photodetection on superconducting qubits in circuit quantum electrodynamics.  Although both measurement schemes give rise to backaction in the form of stochastic phase rotations, which leads to dephasing, we show that this can be perfectly undone provided that the measurement signal is fully accounted for. This result improves upon that of Ref.~\cite{TornbergPRA2010}, showing that the method suggested can be made to realize a perfect two-qubit parity measurement. We propose a benchmarking experiment on a single qubit to demonstrate the method using homodyne detection. By analyzing the limited measurement efficiency of the detector and bandwidth of the amplifier, we show that the parameter values necessary to see the effect are within the limits of existing technology.
\end{abstract}

\pacs{03.67.Pp, 03.65.Yz, 42.50.Dv, 42.50.Pq}

%\keywords{Suggested keywords}
\maketitle

%%%%%%%%%%%%%%%%%%%%%%%%%%%%%%%%%%%

\section{Introduction}

%Recent experimental achievements in measuring \cite{???} and controlling \cite{???} superconducting qubits have opened up for the realization of single-shot qubit read-out, a fundamental criterium to be fulfilled by any physical realization of a quantum computer \cite{DiVincenzo00}. Such a capability would enable \emph{e.g.}~entanglement generation by measurement  \cite{HutchisonCanJofPhys2009, BishopNJP2009} and open up for the implementation of error correction protocols and fault tolerant quantum computing \cite{NielsenChuang}. It is known that measuring a quantum system introduces backaction which can limit the fidelity of the measurement. To realize a high-fidelity measurement, it is therefore necessary to understand the nature of the backaction and how it can be mitigated or, in the best case, avoided. 

Recent developments in the field of superconducting qubits realizing long qubit coherence times \cite{PaikPRL2011}, non-destructive single-shot homodyne measurements \cite{DewesArXiv2011}, use of fast FPGA's \cite{BozyigitNatPhys2010, MartinisQIP2009}, and approaching quantum limited parametric amplifiers \cite{MalletPRL2011}, have opened up for the realization of \emph{e.g.}~entanglement generation by measurement \cite{HutchisonCanJofPhys2009, BishopNJP2009} and the implementation of error correction protocols and fault tolerant quantum computing \cite{NielsenChuang}. It is known that measuring a quantum system introduces backaction which sometimes can limit the fidelity of the measurement. To realize a high-fidelity measurement, it is therefore necessary to understand the nature of the unnecessary backaction and how it can be mitigated or, in the best case, avoided. 

In this paper, we discuss how to completely undo the dephasing induced by measurement on qubits in a circuit quantum electrodynamics (cQED) setup \cite{BlaisPRA2004, WallraffNature2004}. Our analysis is focused on the read-out of a single qubit, but it is equally valid for the case of joint measurement on two qubits, which enables perfect parity measurement in this system. The reason for largely limiting the discussion to the single-qubit case is that the effect can be equally well understood and observed in a one-qubit measurement, which is beneficial from an experimental point of view. 
%Such an experiment would serve as a powerful benchmark to show how well qubits in a cQED architecture can be controlled and measured.

In circuit QED, the qubit is coupled to a microwave resonator which gives rise to a state dependent shift of the resonator frequency. The qubit and the field states become entangled and it is possible to read out the state of the qubit by measuring the state of the emitted microwave field. The homodyne measurement of one and two qubits was analyzed in Refs.~\cite{GambettaPRA2008,KorotkovArxiv} and \cite{LalumierePRA2010}, respectively. In the latter it was shown that when measurement parameters are tuned to realize a parity measurement, there is unwanted  backaction in the form of a stochastic phase causing dephasing in the post measurement state. In Ref.~\cite{TornbergPRA2010} it was shown that this backaction could be partly undone by recording the homodyne current and applying conditional control pulses to undo the phase uncertainty. 

In this work, we extend the analysis of Ref.~\cite{TornbergPRA2010} to show that not only part of the measurement-induced dephasing, but all of it, can be undone provided that the measurement signal is turned off and all photons pumped into the resonator are accounted for. The result is in agreement with the simpler picture of a strong projective measurement of one of the field quadratures for which it is easy to see that the dephasing can be undone given that no knowledge about the initial superposition is obtained.  

To further explore the connection between information gain and dephasing, we consider the backaction when the field is measured using direct photodetection. This is interesting from an experimental point of view due to the recent developments in photon counting devices in the microwave spectrum \cite{ChenPRL2011,PeropadrePRA2011,RomeroPRL2009}. Theorywise, the read-out is interesting because the measurement adds additional backaction apart from the fundamental projection postulated by quantum mechanics \cite{BraginskyKhalili}. By deriving an effective stochastic master equation for the qubit degrees of freedom, we show that, when the photons contain no information about the qubit state, it is possible to completely undo the dephasing by recording the time when each photon is detected. Hence, the connection between information and backaction is established in this case as well. 

Finally, we suggest the implementation of a simple one-qubit experiment to test the undoing of dephasing in the homodyne case and show that it is possible to observe the effect with existing state of the art measurement devices. 

This paper is organized as follows. In \secref{sec_toysystem} we show the connection between information gain and backaction using a simple toy model consisting of two entangled qubits and a projective measurement. In \secref{sec_system} we present the model describing measurement of one qubit in a resonator, considering two cases: homodyne detection and photodetection. In \secref{sec_dephasing} we show, for both detection schemes, how the measurement-induced dephasing can be perfectly undone. For homodyne measurement, the result holds equally well for the case of two qubits. In \secref{sec_experiment} we describe an experiment to test this in circuit QED for the case of homodyne detection, and discuss suitable parameters. Conclusions are given in \secref{sec_conclusion}.

\section{Information gain vs. backaction}\label{sec_toysystem}

To illustrate the connection between information gain and backaction we consider two entangled qubits,
\be\label{eq:state}
\ket{\psi} = \alpha\ket{gg} + \beta\ket{ee}.
\ee
It is clear that making a measurement along the z-axis of qubit two, $\sigma_z^{(2)}$, gives information about  $|\alpha|^2$ and $|\beta|^2$, with the post measurement state of the first qubit given by the mixed state $\rho_\mathrm{qb1} = \mathrm{diag}(|\alpha|^2,|\beta|^2)$. If we, on the other hand, choose to measure $\sigma_x^{(2)}$, we will get the results $\pm1$ with equal probability, hence revealing no information about $\alpha$ and $\beta$. The post measurement state of qubit one is 
\[
\ket{\psi}_\mathrm{qb1} = 
\left\{
\begin{array}{l}
  \alpha\ket{g} + \beta\ket{e}, \qquad \mathrm{if~ } \sigma_x^{(2)} = +1,  \\
  \alpha\ket{g} -  \beta\ket{e}, \qquad \mathrm{if~} \sigma_x^{(2)} = -1  ,
\end{array}
\right .
\]
resulting in a pure state, which is either left untouched or phase flipped. Note that, depending on the measurement result, one can reverse the effect of measurement, leaving the state uncorrupted. 

The situation in cQED is very similar to this model albeit more complex in two senses; the second qubit is replaced with the more complex field state, and the measurement cannot be considered strong but is instead happening on a timescale comparable to the resonator linewidth. In the following we show that the simple symmetry between information gain and backaction is preserved despite these complications.

\section{System}\label{sec_system}

\subsection{Hamiltonian and master equation}

We consider a circuit QED setup with one qubit coupled to a resonator driven by a measurement signal described by the Hamiltonian \cite{BlaisPRA2004}
\begin{eqnarray}
H &=& \frac{\omega_a}{2}\sigma_z + \omega_r a^\dag a + g \left(a^\dag \sigma_- + a\sigma_+ \right) \nonumber\\
&& + \left(\varepsilon_d (t)a^\dag e^{-i\omega_d t} + \varepsilon_d^* (t) a e^{i\omega_d t} \right),
\end{eqnarray}
where we have set $\hbar = 1$, $\omega_a (\omega_r)$ is the resonance frequency of the qubit (resonator), $g$ is the coupling strength between the resonator and the qubit, $a (a^\dag)$ is the annihilation (creation) operator for the resonator, $\sigma_- (\sigma_+)$ is the lowering (raising) operator of the qubit, $\varepsilon_d (t)$ is the amplitude of the measurement drive, and $\omega_d$ is the frequency of the measurement signal. 

In the dispersive regime, $|\Delta| = |\omega_a - \omega_r| \gg g$, the effective Hamiltonian of the system is given by \cite{BlaisPRA2004}
\begin{equation}
H_\text{eff} = \frac{\tilde{\omega}_a}{2}\sigma_z + \Delta_r a^\dag a + \chi a^\dag a\sigma_z + \left(\varepsilon_d (t)a^\dag + \varepsilon_d^* (t) a \right), 
\end{equation}
where $\chi = g^2/\Delta$, $\tilde{\omega}_a = \omega_a + \chi$,  $\Delta_r = \omega_r - \omega_d$, and we have moved to a frame rotating with the measurement signal frequency $\omega_d$. The master equation for the cavity+qubit system in this frame is \cite{CarmichaelBook}
\begin{eqnarray}
\dot{\rho} &=& -i\comm{H_\text{eff}}{\rho} + \kappa\lind{a}\rho + \gamma_1\lind{\sigma_-}\rho \nonumber\\
&&+ \frac{\gamma_\phi}{2} \lind{\sigma_z}\rho, 
\label{UntransfMasterEq}
\end{eqnarray}
where $\rho$ is the density matrix of the system, $\lind{X}\rho = X\rho X^\dag - \frac{1}{2} X^\dag X\rho - \frac{1}{2} \rho X^\dag X$ are Lindblad operators \cite{Lindblad1976}, $\kappa$ is the resonator decay rate and $\gamma_1$, $\gamma_\phi$ are the qubit relaxation and dephasing rates respectively. If the resonator state is coherent, and $\kappa \gg \gamma_1$, the field amplitudes solving \eqref{UntransfMasterEq} are given by \cite{GambettaPRA2006}
\begin{eqnarray}\label{eq_mastertot}
\dot{\alpha}_g(t) &=& - i \varepsilon_d(t) - i \left(\Delta_r - \chi \right)\alpha_g(t) - \frac{\kappa}{2}\alpha_g(t), \label{AlphaGDot}\\
\dot{\alpha}_e(t) &=& - i \varepsilon_d(t) - i \left(\Delta_r + \chi \right)\alpha_e(t) - \frac{\kappa}{2}\alpha_e(t),\label{AlphaEDot}
\end{eqnarray}
depending on whether the qubit is in the ground state $\ket{g}$ or the excited state $\ket{e}$. 

\subsection{Stochastic master equations}

Although the master equation in \eqref{UntransfMasterEq} describes the evolution of the open system, we need to take into account the conditional evolution given one certain measurement trace. In this section, we consider two types of measurement; homodyne detection and photodetection.

The stochastic master equation (SME) \cite{Wiseman1993} describing the system evolution conditioned on homodyne detection is given by
\begin{equation}
\rd\rho = \mathcal{L}_{\text{tot}}\rho\rd t + \sqrt{\kappa\eta}\meas{ae^{-i\phi}}\rho\rd W(t),
\label{UntransfSMEHom}
\end{equation}
where $\mathcal{L}_{\text{tot}}\rho$ is the RHS of \eqref{UntransfMasterEq}, $\eta$ is the measurement efficiency and $\meas{c}\rho = c\rho + \rho c^\dag - \expec{c + c^\dag}\rho$ is the measurement operator with $\expec{A} = \tr{A\rho}$.  The phase of the local oscillator is given by  $\phi$, and $\rd W(t)$ is a Wiener increment defined by $\expecc{\rd W(t)} = 0$ and $\expecc{\rd W(t)^2} = \rd t$. The homodyne current is given by
\begin{equation}
j(t)\rd t = \sqrt{\kappa\eta}\expec{a e^{-i\phi} + a^\dag e^{i\phi}}\rd t + \rd W(t).
\end{equation}
%

%\begin{figure}\centering
%\includegraphics[width=\linewidth]{Figures/HomodyneSetup.eps}
%\caption[]{The setup for homodyne detection.} \label{HomodyneSetup}
%\end{figure}

For the case of photodetection the SME is \cite{QuantumMeasurement}
\begin{eqnarray}
\rd\rho &=& -\frac{i}{\hbar}\comm{H_\text{eff}}{\rho}\rd t + (1-\eta)\kappa\lind{a}\rho\rd t \nonumber\\
&& + \gamma_1\lind{\sigma_-}\rho\rd t + \frac{\gamma_\phi}{2} \lind{\sigma_z}\rho \rd t \nonumber\\
&& + \measg{a}\rho\rd N(t) - \frac{1}{2}\eta\kappa\meas{a^\dag a}\rho \rd t,
\label{UntransfSMEphoto}
\end{eqnarray}
where $\measg{c}\rho = \frac{c\rho c^\dag}{\expec{c^\dag c}} - \rho$, and $\rd N(t)$ is a Poisson point process with the properties $\rd N(t)^2 = \rd N(t)$ and $\expecc{\rd N(t)} = \eta\kappa\expec{a^\dag a}\rd t$.

Although \eqref{UntransfSMEHom} and \eqref{UntransfSMEphoto} can be used directly to study the measurement backaction, they provide little intuition and understanding. Instead we choose to trace out the resonator degrees of freedom and work with effective equations for the qubit degrees of freedom only.

\subsection{Effective stochastic master equations}

The effective SME for the qubit degrees of freedom when read out through homodyne detection was derived in Ref.~\cite{GambettaPRA2008} and is given by
\begin{eqnarray}
\rd\rhoq &=& -i\frac{\tilde{\omega}_a + B}{2}\comm{\sigma_z}{\rhoq}\rd t + \gamma_1\lind{\sigma_-}\rhoq\rd t \nonumber\\
&&+ \frac{\gamma_\phi + \Gamma_d}{2}\lind{\sigma_z}\rhoq \rd t \nonumber\\
&&+ \sqrt{\kappa\eta} \meas{\Pi_\alpha e^{-i\phi}}\rhoq\rd W(t),
\label{EffectiveSMEHom}
\end{eqnarray}
where $\Pi_\alpha = \alpha_g\Pi_g + \alpha_e\Pi_e$. Here $\Gamma_d = 2\chi\im\left(\alpha_g\alpha_e^*\right)$ is the measurement-induced dephasing and $B = 2\chi\re\left(\alpha_g\alpha_e^*\right)$ gives the AC Stark shift. The homodyne current is given in terms of qubit operators by
\begin{eqnarray}
j(t)\rd t =  \sqrt{\kappa\eta}\expec{\Pi_\alpha e^{-i\phi} + \Pi_\alpha^\dag e^{i\phi}}_{\rhoq}\rd t + \rd W(t).
\label{EffectiveHomodyneCurrent}
\end{eqnarray}
In the Appendix, we follow the procedure in Ref.~\cite{GambettaPRA2008} to derive an effective SME for the case of photodetection.  Applying the displacement transformation
\begin{equation}
\bP = \Pi_g D\left[\alpha_g\right] + \Pi_e D\left[\alpha_e\right],
\label{DisplacementTransf}
\end{equation}
where $\Pi_{e/g} = \ket{e/g}\bra{e/g}$ and $ D\left[\alpha\right] = \exp\left(\alpha a^\dag - \alpha^*a \right)$ on \eqref{UntransfSMEphoto} takes the system to a frame of reference where the resonator field is in the vacuum state. This makes it possible to trace out the resonator degrees of freedom, enabling us to derive the effective SME for the qubit conditioned on photodetection 
\begin{eqnarray}\label{EffectiveSMEPhoto}
\rd\rhoq &=& -i\frac{\tilde{\omega}_a + B}{2}\comm{\sigma_z}{\rhoq}\rd t + \gamma_1\lind{\sigma_-}\rhoq\rd t \nonumber\\
&& + \frac{\gamma_\phi + \Gamma_d}{2}\lind{\sigma_z}\rhoq \rd t  - \eta\kappa\lind{\Pi_\alpha}\rhoq\rd t \\
&& - \frac{1}{2}\eta\kappa\meas{\Pi_\alpha^\dag\Pi_\alpha}\rhoq\rd t + \measg{\Pi_\alpha}\rhoq\rd N(t) \nonumber.
\end{eqnarray}
%

%%%%%%%%%%%%%%%%%%%%%%%%%%%%%%%%%%%

\section{Dephasing}\label{sec_dephasing}

To study the unwanted measurement-induced dephasing in the single qubit case, we consider the case $\Delta_r = 0$, which implies $\alpha_g = -\alpha_e^\ast$
%$\im\left(\alpha_g\right) = \im\left(\alpha_e\right)$, $\re\left(\alpha_g\right) = -\re\left(\alpha_e\right)$ and $\abs{\alpha_g} = \abs{\alpha_e}$ 
for the field state. In this case, doing homodyne detection with $\phi = \pi/2$ (measuring the projection of $\alpha_i$ on the imaginary axis) or photodetection (measuring the photon number) will not give any information about the qubit state. However, the measurement still gives rise to dephasing in the form of stochastic phase rotations. For the homodyne case, the analysis is equally valid for a measurement of two-qubit parity where the unwanted backaction gives rise to dephasing within the negative parity subspace \cite{TornbergPRA2010}. In analogy with \secref{sec_toysystem}, we show how this can be perfectly undone given the measurement result. 

\subsection{Homodyne detection}\label{subsec_Dephasing_Homodyne}

We consider the effective SME, \eqref{EffectiveSMEHom}, with $\gamma_1 = \gamma_\phi = 0$, and exclude the coherent evolution since this only gives rise to a deterministic phase which can be undone. In this case the SME can be solved analytically with the solution 
%using that $\meas{\Pi_\alpha e^{-i\pi/2}}\rho = i \re\left(\alpha_g \right)\comm{\sigma_z}{\rho}$, and dropping from here on the superscript $qb$:
%
%\begin{eqnarray}
%\rd\rho &=&  \frac{\Gamma_d}{2}\lind{\sigma_z}\rho \rd t + i \re\left(\alpha_g \right)\sqrt{\kappa\eta} \comm{\sigma_z}{\rho}\rd W(t).
%\label{EffectiveSMEHomDeph}
%\end{eqnarray}
%
%This gives the following two equations for the elements of the density matrix:
%
%\begin{eqnarray}
%\rd\rho_{gg} &=& 0, \\
%\rd\rho_{eg} &=& -\Gamma_d\rho_{eg} \rd t + 2i\re\left(\alpha_g \right)\sqrt{\kappa\eta}\rho_{eg}\rd W(t)
%\end{eqnarray}
%
%The first equation tells us that the measurement does not affect the qubit population. The second equation has the analytic solution
%
\begin{eqnarray}
\rho_{gg}(t) &=& \rho_{gg}(0), \nonumber \\
\rho_{eg}(t) &=& \rho_{eg}(0)\exp\bigg(\int_0^t \left(-\Gamma_d(s) + 2\kappa\eta \re^2\left(\alpha_g(s) \right) \right)\id s \nonumber\\
&&+ 2i\sqrt{\kappa\eta}\int_0^t \re\left(\alpha_g(s) \right) \rd W(s) \bigg),
\label{RhoEGAnalyticSolutionHom}
\end{eqnarray}
where, from now on, we drop the superscript "qb". The second equation gives the dephasing and was investigated in  Ref.~\cite{TornbergPRA2010} for the two-qubit case in the context of parity measurement. There, a constant measurement tone was assumed and it was found that there was inevitable dephasing resulting in a mixed post-measurement state. This was attributed to the finite response time of the cavity. Here, we extend the analysis and turn off the measurement signal at time $t_\text{meas}$ such that 
%We investigate the real part of the exponent in this expression further, assuming a perfect measurement device ($\eta = 1$). In the case of a constant measurement signal amplitude, turned on at time $t=0$ (\emph{i.e.} $\epsilon_d(t) =  \epsilon_m\Theta(t)$, where $\Theta(t)$ is the Heaviside step function), solving \eqref{AlphaGDot} gives
%
%\begin{eqnarray}
%\alpha_g(t) = \frac{2\epsilon_m}{2\chi + i\kappa}\left(1 - \exp\left(-\left(\frac{\kappa}{2} - i\chi \right) t \right)\right)\Theta(t)
%\end{eqnarray}
%
%and from this follows
%
%\begin{eqnarray}
%&&\lim_{t \rightarrow \infty} \int_0^t \left(-\Gamma_d(s) + 2\kappa\left(\re\left(\alpha_g(s) \right)\right)^2 \right)\id s \nonumber\\
%&=& - \frac{32\epsilon_m^2\chi^2}{\left(\kappa^2 + 4\chi^2\right)^2}. 
%\label{LimitConstantMeasSignal}
%\end{eqnarray}
%
%However, if the measurement signal is turned off at some time $t_\text{meas}$, \emph{i.e.} 
$\epsilon_d(t) =  \epsilon_m\left(\Theta(t) - \Theta(t-t_\text{meas})\right)$, where $\Theta(t)$ is the Heaviside step function. In this case the solution to  \eqref{AlphaGDot} is given by 
\begin{eqnarray}
\alpha_g(t) &=& \frac{2\epsilon_m}{2\chi + i\kappa}\bigg[\left(1 - \exp\left(-\left(\frac{\kappa}{2} - i\chi \right) t \right)\right)\Theta(t) \\
- \bigg(1 &-& \exp\left(-\left(\frac{\kappa}{2} - i\chi \right) \left(t-t_\text{meas}\right)\right)\bigg)\Theta(t-t_\text{meas}) \bigg], \nonumber
\end{eqnarray}
which inserted in the real part of the exponent in \eqref{RhoEGAnalyticSolutionHom} gives
\begin{eqnarray}\label{eq_zerodephasing}
&\lim_{t \rightarrow \infty}& \int_0^t \left(-\Gamma_d(s) + 2\kappa\re^2\left(\alpha_g(s) \right) \right)\id s \nonumber \\
=&\lim_{t \rightarrow \infty}&\frac{(\alpha_g(t) - \alpha_e(t))^2}{4} = 0,
\end{eqnarray}
where it should be noted that we have assumed a quantum limited measurement $\eta = 1$.
Hence, by turning off the measurement signal and allowing for all the information that is encoded about the qubit state to be measured, we completely recover the initial pure state of the qubit except for the stochastic phase seen in the second row of \eqref{RhoEGAnalyticSolutionHom}. This is one of the main results of this paper.  If, on the other hand, the measurement signal isn't turned off, there will always be some photons left in the resonator and the information which is thus withheld from us gives the residual dephasing that was found in Ref.~\cite{TornbergPRA2010}. 

The fact that the measurement induced dephasing is zero when not acquiring information about the qubit state is not apparent when looking at \eqref{EffectiveSMEHom}. However, since the decrease of the off diagonal element is associated with the gradual projection onto one of the measurement eigenstates, the result is expected. This was also found in \cite{KorotkovArxiv} using a quantum bayesian formalism in the limit $\chi \ll \kappa$.

The imaginary part of the exponent in \eqref{RhoEGAnalyticSolutionHom} gives stochastic phase kicks to the qubit. Averaging this over many trajectories would give rise to dephasing. However, since we measure the homodyne current, which simplifies from \eqref{EffectiveHomodyneCurrent} to 
\begin{eqnarray}
j(t)\rd t = 2 \sqrt{\kappa\eta}\im\left(\alpha_g(t)\right)\rd t + \rd W(t)
\label{EffectiveHomodyneCurrentPi2}
\end{eqnarray}
for the case $\phi = \pi/2$, and we know $\alpha_g(t)$, we can extract $\rd W(t)$ and the stochastic phase kicks from our measurement. In conclusion, if $\eta = 1$ we are able to completely determine the measurement-induced phase kick to the qubit and undo it after each run of the experiment, essentially undoing the measurement-induced dephasing. The elimination of the stochastic phase was also discussed in \cite{KorotkovArxiv}. Here, a direct feedback controller was used to modulate the qubit frequency. This suggestion requires the modulation to be done on the time-scale of the fluctuations in the measured current. Our suggestion is not limited by this since all the phase is undone using a single control pulse at the end of the measurement. 

The connection between phase rotation and measurement result can be made intuitively clear by considering the strong version of the measurement, \emph{i.e}.~a projection on one of the eigenstates  of the field quadrature $P = i(\dagg{a} - a)/\sqrt{2}$. The post measurement state of the qubit is given by 
\be
\ket{\psi}_\mathrm{qb} = \frac{1}{N}\left(\delta \langle p|\alpha_g \rangle \ket{g} + \gamma \langle p|\alpha_e\rangle \ket{e}\right),
\ee
where $N$ is a normalization and the overlap integral $ \langle p| \beta\rangle$ is given by \cite{ZollerGardiner}
\be\label{eq_overlap}
\langle p| \beta\rangle = \frac{1}{\pi^{1/4}} \exp\left[ -\im(\beta)^2 -2\beta^2 - (p + i \sqrt{2}\beta)^2 \right],
\ee
and $\delta$ and $\gamma$ are arbitrary complex numbers satisfying $|\delta|^2+|\gamma|^2 = 1$ and $P\ket{p} = p\ket{p}$. The off-diagonal element in the post measurement density matrix is hence given by 
\be
\rho_{eg}^\text{post} = \delta^\ast \gamma e^{-i 2\sqrt{2} \re(\alpha_g)p},
\ee
%
%\be
%\ket{\psi}_\mathrm{qb} \propto \delta e^{i\phi(p)}\ket{g} + \gamma e^{-i\phi(p)}\ket{e},
%\ee
%
%where 
%
%\be
%\phi(p) = 2\re(\alpha_g)p + 4\re(\alpha_g)\im(\alpha_g).
%\ee
%
where we have used the specific symmetries of the measurement setup. In the strong measurement limit, the only effect on the state is to apply a stochastic phase rotation between $\ket{g}$ and $\ket{e}$ whose magnitude depends on the measurement outcome $p$. This is in analogy with the analysis above. The difference between the strong and weak measurement only lies in the fact that the entire measurement trace must be accounted for if the measurement is weak. 
%In \secref{sec_experiment}, we take into account the limited bandwidth of the measurement apparatus and show that it is still possible to distill the effect of phase kick-back using state of the art amplifiers.  
%

\begin{figure}
    \centering{
    \subfigure
    {
	\includegraphics[width=\linewidth]{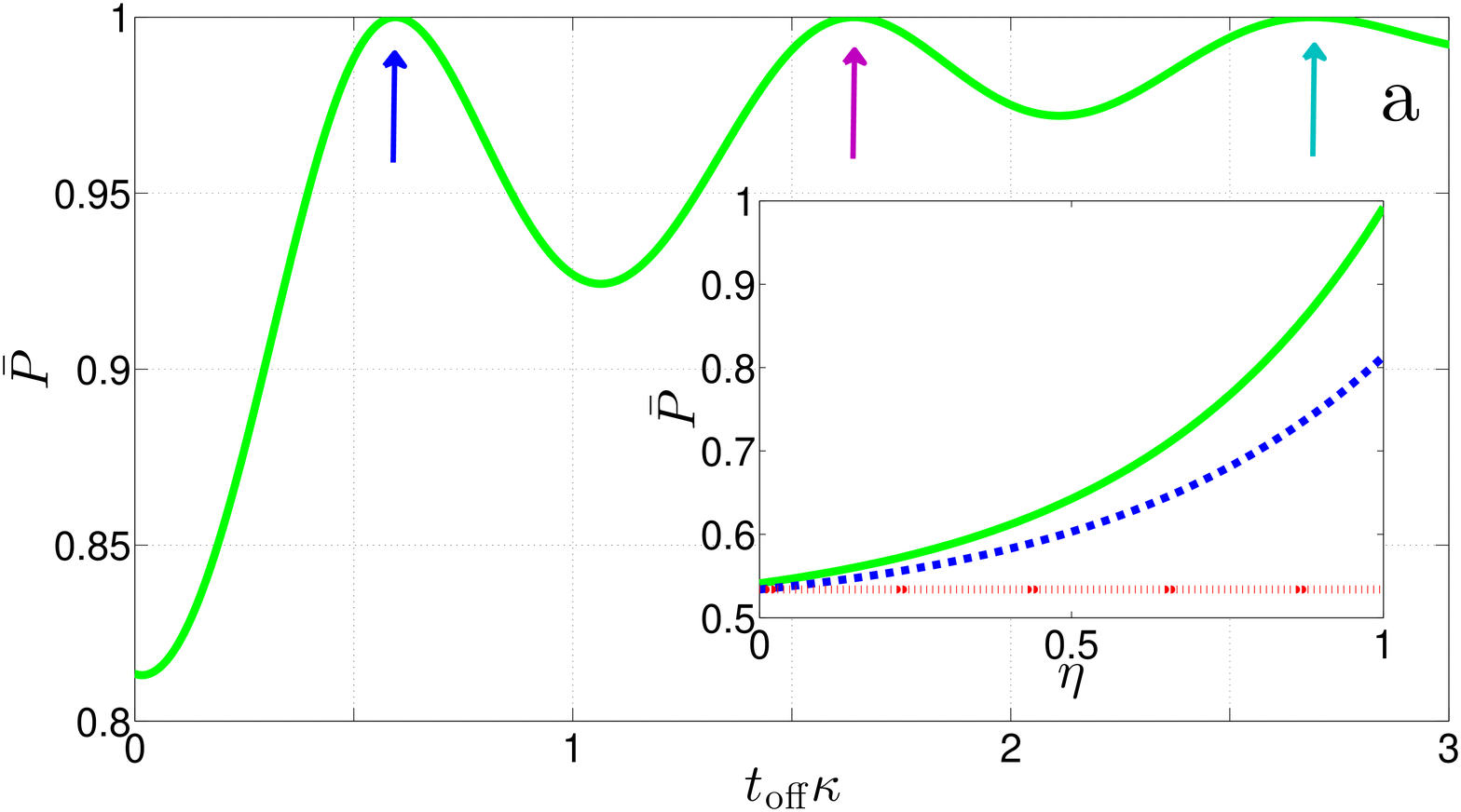}
	 \label{PurityTime_a}
    }    
    \subfigure
    {
	\includegraphics[width=\linewidth]{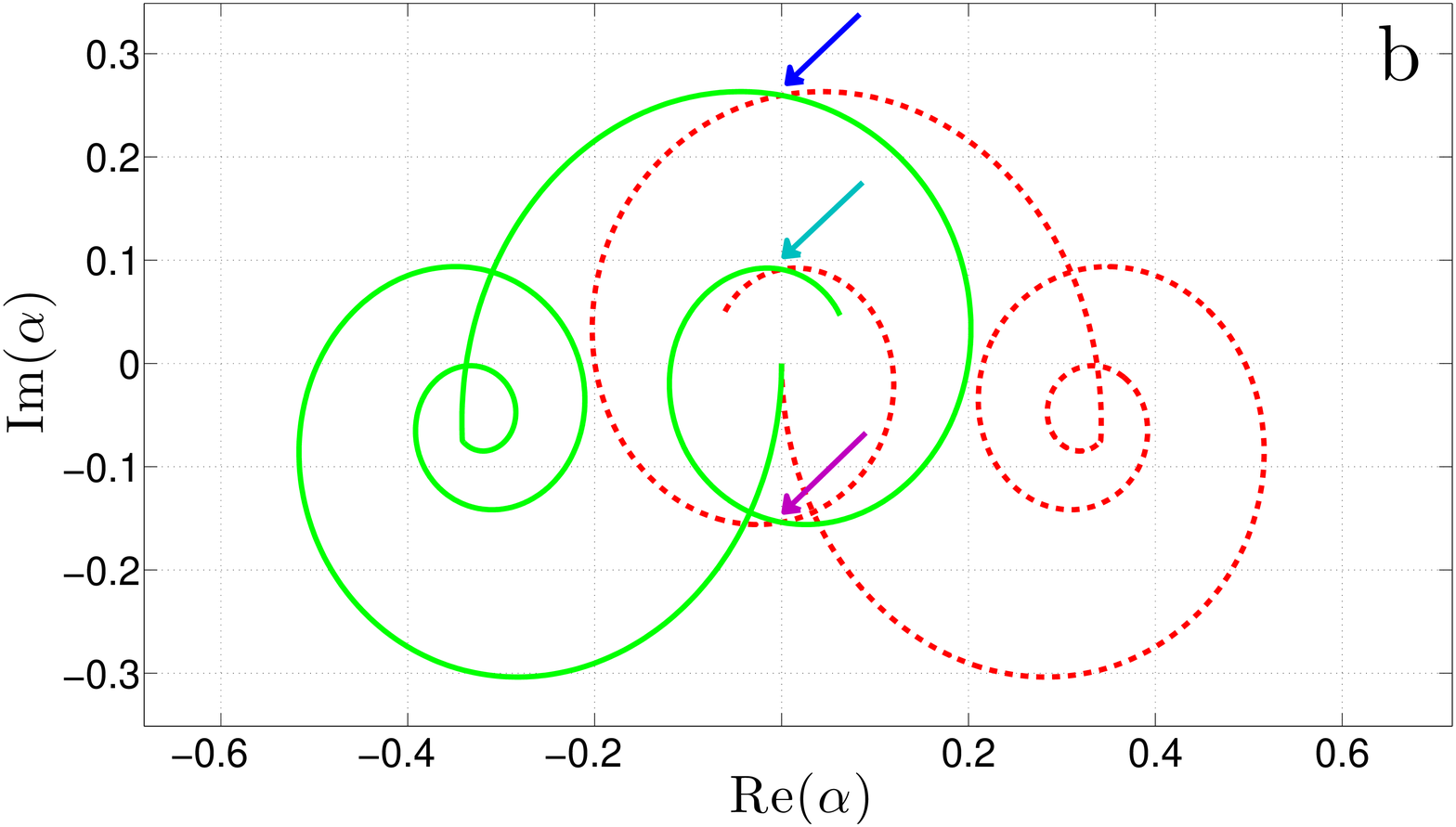}
	 \label{PurityTime_b}
    }
    }
    \caption{(Color online) {\bf a}) The purity as a function of the waiting time $t_\text{off}$.  After each measurement, a phase correction conditioned on the measurement trace was applied as described in the text. The parameters are given by $\varepsilon_d = \kappa $, $t_\text{meas} = 5/\kappa$ and $\chi = 3\kappa$. The inset shows the purity as a function of measurement efficiency for the case of; no feedback (red, dotted), feedback without accounting for all photons (blue, dashed) and feedback with full accounting (solid green). Here, $t_\text{off} = 3/\kappa$.\\
 {\bf b}) The time evolution of the cavity field $\alpha_e$ (green) and $\alpha_g$ (red, dashed). After the measurement signal is turned off, the field starts to spiral back towards the origin. At certain points in time (arrows) $\alpha_e = \alpha_g$, for which the qubit state is pure.}
    	 \label{PurityTime}
\end{figure}

In \figref{PurityTime_a} we plot the purity as a function of the waiting time after the measurement signal has been turned off, $t_\text{off}$. Using the procedure outlined above to correct the measurement-induced phase kicks, we see that the post measurement state approaches a pure state when all the photons have been accounted for. Observe that the purity oscillates during the measurement such that the qubit state is pure at certain points in time. We can understand this by looking at the time evolution of the resonator states which is plotted in \figref{PurityTime_b}.  After the measurement signal has been turned off $\alpha_g$ and $\alpha_e$ spiral towards the origin. We see that the times for which $\bar{P} = 1$ coincides with the times at which the resonator states overlap. The overlap integral in \eqref{eq_overlap} is the same at these points such that the photons remaining in the resonator carry no information about the qubit state. To ensure that the purity remains 1, however, we must wait until all photons have left the resonator, which takes a few resonator lifetimes.

\subsubsection*{Parity measurement}
The result presented in \eqref{eq_zerodephasing} can easily be extended to the case of joint parity measurement on two qubits which was studied in detail in Ref.~\cite{TornbergPRA2010}. In this case, the qubit-resonator couplings and detunings are given by $g_1 = g_2$ and $\Delta_1 = - \Delta_2$ which enables an observer to distinguish the parity of a two-qubit state. As shown in \cite{LalumierePRA2010}, this results in dephasing within the negative parity subspace spanned by $\ket{eg}$ and $\ket{ge}$. In analogy with the single-qubit measurement, this dephasing originates from the entanglement with the field states $\alpha_{ge}$ and $\alpha_{eg}$.  Since these evolve in the same way as $\alpha_{g}$ and $\alpha_{e}$ (but with the replacement $\chi \to 2\chi$ in Eqs.~(\ref{AlphaGDot}) and (\ref{AlphaEDot})), the result in \eqref{eq_zerodephasing} holds equally well with the substitution $\alpha_g \to\alpha_{ge}$ and $\alpha_e \to \alpha_{eg}$, which shows that the measurement-induced dephasing can be completely undone in the two-qubit case as well. In this way it is possible to realize a perfect parity measurement in cQED.  

%\subsubsection{Two qubits - parity measurement}

%In Ref.~\cite{TornbergPRA2010} our scheme for undoing measurement-induced dephasing, except for turning off the measurement signal, was discussed for the case of homodyne detection in a parity measurement on two qubits.  

%MORE DETAILED DESCRIPTION??

%In \figref{TwoQubitsPFC} we plot the purity $\bar{P}$, the concurrence $\bar{C} = C\left(\expecc{\rho}\right)$, and the fidelity
%
%\begin{equation}
%\bar{F} = \frac{n_+\brakket{\psi_+}{\expecc{\rho}}{\psi_+} + n_-\brakket{\psi_-}{\expecc{\rho_}}{\psi_-}}{n_+ + n_-},
%\end{equation}
%
%where $n_+$($n_-$) is the number of measurements giving results +(-) and $\expecc{\rho_+}$ ($\expecc{\rho_-}$) is the average over the respective density matrices. We see that waiting some time after turning off the measurement signal gives excellent results.

%\begin{figure}\centering
%\includegraphics[width=\linewidth]{Figures/PFC_vert_110908.eps}
%\caption[]{The purity, fidelity and concurrence for two qubits after measurement, with no phase correction (red dots), with phase correction after $t=t_\text{meas}$ (blue dots), and with phase correction after $t=t_\text{meas} +t_\text{off} $ (green dots). Parameters the same as for Figs. 3-5 in Ref.~\cite{TornbergPRA2010} with $t_\text{off}  = 3/\kappa$. \label{TwoQubitsPFC}}
%\end{figure}

%%%%%%%%%%%%%%%%%%%%%%%%%%%%%%%%%%%

\subsection{Photodetection}

In \secref{subsec_Dephasing_Homodyne}, we saw that it is possible to gain insight about the backaction of the weak measurement by looking at the corresponding strong measurement limit. In the case of photodetection the detector adds additional backaction to the system by absorbing the photons that are detected. Hence, the post-measurement state is vacuum and not an eigenstate of the photon number operator corresponding to the number of measured photons. For the choice of parameters considered in \secref{subsec_Dephasing_Homodyne} there is however nothing in the field states which allows us to distinguish the qubit states when measuring photon number. We now show that, despite the additional backaction, the purity of the qubit state is preserved as it should be if no information is acquired. To this end we consider the effective photodetection SME, \eqref{EffectiveSMEPhoto}, with the same choice of parameters as for the homodyne case
\begin{eqnarray}
\rd\rho &=& \frac{\Gamma_d}{2}\lind{\sigma_z}\rho \rd t - \eta\kappa\lind{\Pi_\alpha}\rho\rd t \nonumber\\
&& - \frac{1}{2}\eta\kappa\meas{\Pi_\alpha^\dag \Pi_\alpha}\rho\rd t + \measg{\Pi_\alpha}\rho\rd N(t).
\label{EffectiveSMEPhotoDeph}
\end{eqnarray}
This gives the following two equations for the elements of the density matrix:
\begin{eqnarray}
\rd\rho_{gg} &=& 0, \\
\rd\rho_{eg} &=& \left(-\Gamma_d + \eta\kappa\left(\alpha_e^2 + \Na \right)\right)\rho_{eg} \rd t \nonumber\\
&&- \frac{\alpha_e^2 + \Na}{\Na}\rho_{eg} \rd N(t),
\end{eqnarray}
where we have introduced the average photon number $\Na \equiv \abs{\alpha_g}^2$. The second equation describes the dephasing and has the analytic solution \cite{PlatenJump}
%
%\begin{eqnarray}
%\rho_{eg}(t) &=& \exp\left(\int_0^t \left(-\Gamma_d(s) + \eta\kappa\left(\alpha_e^2(s) + \Na(s) \right) %\right)\id s\right) \nonumber\\
%&&\times \rho_{eg}(0)\prod_{t_N\leq t}-\frac{\alpha_e^2(t_N)}{\Na(t_N)}
%\label{RhoEGAnalyticSolutionPhoto}
%\end{eqnarray}
\begin{eqnarray}
\rho_{eg}(t) &=& \exp\left(\int_0^t \left(-\Gamma_d(s) + \eta\kappa\left(\alpha_e^2(s) + \Na(s) \right) \right)\id s\right) \nonumber\\
&&\times \rho_{eg}(0)\prod_{t_N\leq t}-e^{i2\xi(t_N)},
\label{RhoEGAnalyticSolutionPhoto}
\end{eqnarray}
where $\xi = \text{Arg}(\alpha_e)$ and $t_N$ are the times at which the photons are detected. 
Observing that 
\begin{eqnarray}
\alpha_e^2 = 2\re^2\left(\alpha_g\right) - \Na + i \im\left(\alpha_e^2\right),
\end{eqnarray}
\eqref{RhoEGAnalyticSolutionPhoto} reduces to (in the case $\eta=1$ and $t\gg t_{\text{meas}}$)
\begin{eqnarray}
\rho_{eg}(t) &=& \rho_{eg}(0)\exp\left(i \int_0^t \kappa \im(\alpha_e^2(s)) \id s\right)\prod_{t_N\leq t}-e^{i2\xi(t_N)},  \nonumber \\
%&&\times \rho_{eg}(0)\prod_{t_N\leq t}-e^{i2\xi(t_N)},
\label{RhoEGAnalyticSolutionPhotoSimplified}
\end{eqnarray}
where we have used the result from \eqref{eq_zerodephasing}. We see that the qubit state acquires a phase kick at each time a photon is detected. The size of this is just twice the phase of the applied read out field which is a deterministic quantity. As for the homodyne case, the measurement backaction preserves the purity of the state and the initial state can be perfectly recovered by an observer who keeps track of the times that the photons arrive at the detector.  

%The imaginary part of the exponent gives a phase contribution to the qubit. The last term in \eqref{RhoEGAnalyticSolutionPhoto} gives a stochastic phase kick at each time $t_N$ a photon is detected. All these phase contributions are known since $\alpha_e(t)$ is known and we know all the $t_N$. This lets us conclude that just as in the case of homodyne detection we can undo the measurement-induced dephasing provided that $\eta = 1$.

%%%%%%%%%%%%%%%%%%%%%%%%%%%%%%%%%%%

\section{Experiment}\label{sec_experiment}

Although experimental progress in microwave photodetectors is promising \cite{ChenPRL2011,PeropadrePRA2011,RomeroPRL2009}, it is, as of today, difficult to test \eqref{RhoEGAnalyticSolutionPhotoSimplified} experimentally. Homodyne detection, on the other hand, is routinely used to read out the single and multi-qubit states in cQED \cite{DiCarloNature2011}, suggesting that, at least, it should be possible to undo the measurement-induced dephasing with this setup. The feedback scheme suggested above assumes a perfect record of the measurement current needed to calculate the necessary conditional phase kick back. Unfortunately, this requires a quantum limited ($\eta = 1$) detector with infinite bandwidth (in order to perfectly resolve $j(t)$). Current state of the art measurements in cQED are done with parametric amplifiers \cite{CastellanosNatPhys2008, CastellanosIEEE2009, MalletPRL2011} having a typical measurement efficiency of $\eta = 0.4$ and bandwidth $\text{BW} = \unit[10]{MHz}$.
%As is clear from \figref{EtaDependence}, this $\eta$ diminishes the observable effect, but there should still be a clear difference between doing feedback or not on the qubit.

\begin{figure}%\centering
\includegraphics[width=\linewidth]{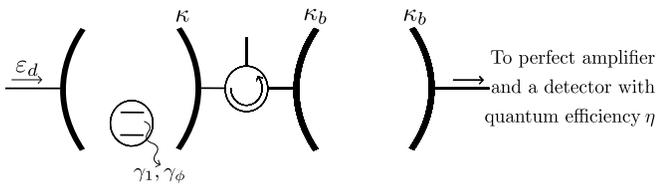}
\caption[]{The setup used to model finite bandwidth of the detector.} \label{2Resonators}
\end{figure}

\begin{figure}%\centering
\includegraphics[width=\linewidth]{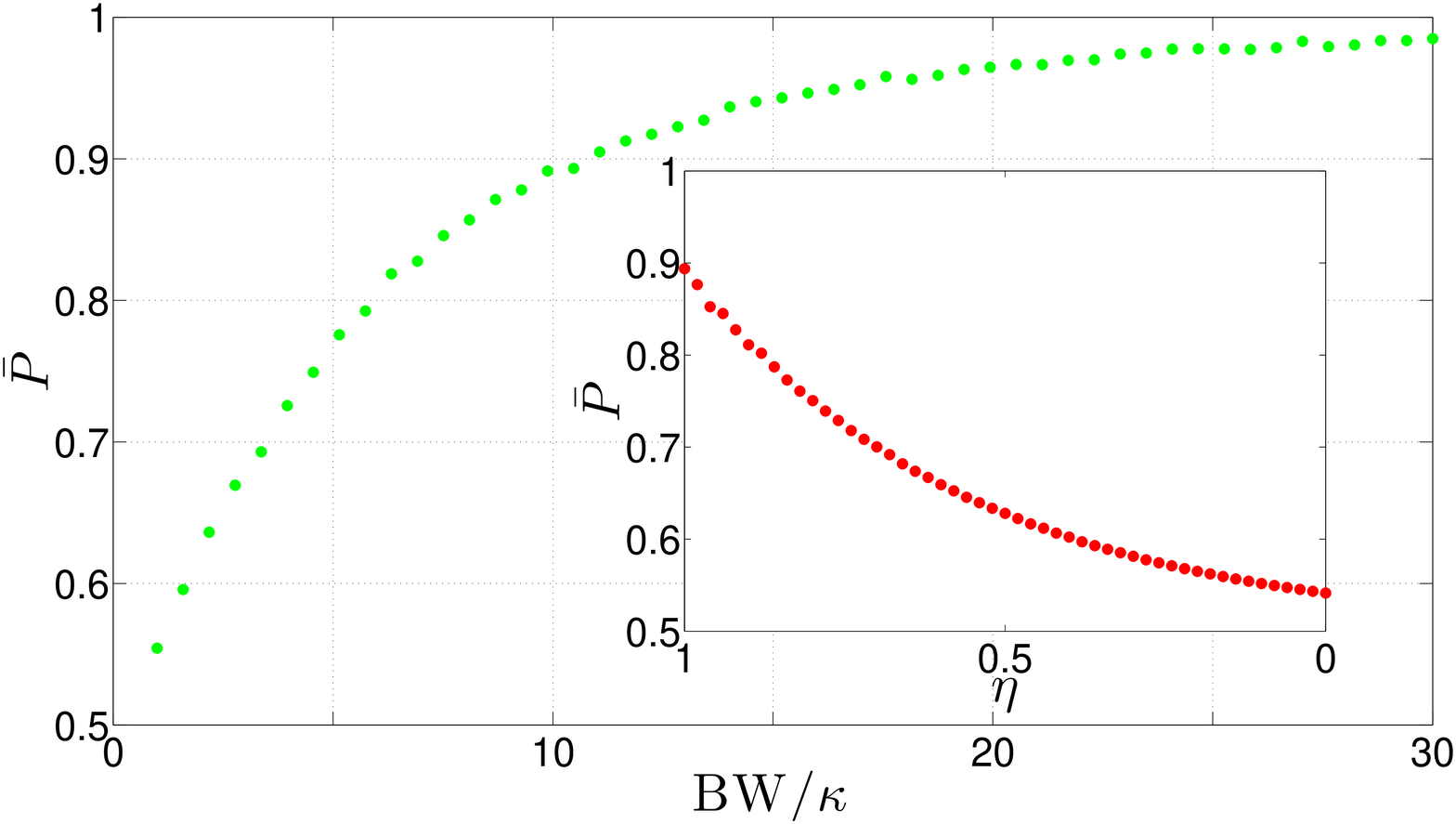}
\caption[]{(Color online) The purity of the qubit density matrix after measurement, with phase correction after $t=t_\text{meas} +t_\text{off} $, as a function of amplifier bandwidth. Parameters are as in \figref{PurityTime} with $t_\text{off} = 10/\kappa$. The inset shows the purity vs measurement efficiency for $\text{BW} = 10\kappa$.} 
\label{PurityBW}
\end{figure}

To see if the effect of feedback can be observed with realistic measurement parameters, we model the effect of finite amplifier BW by inserting a second resonator between the system of interest and an amplifier having infinite bandwidth and detector with measurement efficiency $\eta$. The second resonator is characterized by a loss rate $\kappa_b = \text{BW}/2$ through each of its sides and annihilation operator $b$. The setup is depicted in \figref{2Resonators}. Using the (S,L,H) formalism for cascaded quantum systems \cite{Gough1,Gough2} the SME for the full system is given by
\begin{eqnarray}\label{eq_fullSME}
\rd\rho &=& \Big[-\frac{i}{\hbar}\comm{H_{\text{eff}}}{\rho} + \kappa\lind{a}\rho + 2\kappa_b\lind{b}\rho \nonumber\\
&&+\gamma_1\lind{\sigma_-}\rho + \frac{\gamma_\phi}{2} \lind{\sigma_z}\rho \nonumber\\
&&- \sqrt{\kappa_{b}\kappa}\left(\comm{\rho a^\dag}{b} + \comm{b^\dag}{a\rho}\right)\Big]\rd t\nonumber \\
&&+\sqrt{\kappa_{b}\eta}\meas{be^{-i\phi}}\rho\rd W(t),
\end{eqnarray}
where, in this case, the homodyne current is given by
\begin{eqnarray}\label{eq_truecurrent}
j(t) \rd t &=& \sqrt{\kappa_{b}\eta}\expec{be^{-i\phi} + b^\dag e^{i\phi}} \rd t + \rd W(t).
\end{eqnarray}
In \figref{PurityBW} we plot the purity of the post-measurement state as a function of the amplifier bandwidth. The results are obtained by numerical integration of \eqref{eq_fullSME} including both resonator's degrees of freedom. Given the homodyne current, we extract the noise process by assuming \eqref{EffectiveHomodyneCurrentPi2}. Since the true current is given by \eqref{eq_truecurrent}, this will lead to a wrong estimation of the phase kick back which limits the fidelity of the process. For the parameter values quoted above, there is however a clear distinction between the purity of the post-measurement state with and without the feedback protocol proposed, thereby suggesting that it should be possible to see this effect using state of the art cQED read-out and control technology. Such an experiment would serve as a powerful benchmark to show how well qubits in a cQED architecture can be controlled and measured.

%We see that we should have at least $\text{BW} > 10\kappa$ to minimize the effect from the finite BW. However, we have to weigh this against the condition $\kappa \gg \gamma_1$.

%\begin{figure}\centering
%\includegraphics[width=\linewidth]{Figures/VaryEtaAnalytic120126.eps}
%\caption[]{The purity of the qubit density matrix after measurement, with no phase correction (red dots), with phase correction after $t=t_\text{meas}$ (blue dots), and with phase correction after $t=t_\text{meas} +t_\text{off} $ (green dots). Parameters are as in \figref{PurityTime} with $t_\text{off} = 3/\kappa$. \label{EtaDependence}}
%\end{figure}

%We believe that it should be possible to test the undoing of measurement-induced dephasing for one qubit with homodyne detection using existing circuit QED technology. Below we outline a possible experimental protocol and investigate how different experimental parameters will affect the result.

%\subsection{Protocol}

Apart from the measurement infidelities in the experimental setup, we have to take into account the time it takes to analyze the measurement result needed to calculate the right kick back. Comparing the coherence times of transmon type qubits ($\sim50\mu s$ \cite{PaikPRL2011}) with current performance of state of the art FPGA's \cite{BozyigitNatPhys2010}, this might be possible but comes at the cost of difficult implementation. Instead of the direct scheme proposed in \secref{subsec_Dephasing_Homodyne}, we therefore propose the following, simple protocol which avoids these complications
\begin{enumerate}
\item Prepare the qubit in the superposition state $\ket{\Psi} = (\ket{0} + \ket{1})/\sqrt{2}$.

\item Apply a weak ``measurement" signal with local oscillator phase $\phi = \pi/2$ during time $t_\text{meas}$ and wait a time $t_\text{off}$. Record the corresponding homodyne current and store the result. This will only induce a stochastic phase rotation which we seek to undo.   

\item Apply a pulse that changes the phase of the qubit randomly by $\phi_0$.

\item Do a strong measurement of the qubit ($\phi = \pi/2$). This does not cause additional phase rotations but projects the qubit state on the measurement eigenbasis. Store the result. 
\end{enumerate}

%(quantum state tomography \cite{LiuPRB2005,SteffenPRL2006,HouckNature2007}).

When this protocol has been repeated many times there is plenty of time to calculate the required phase $\phi_\text{calc}$ for each measurement run. Given this, post-select the runs where $\phi_0$ is in the range $[\phi_\text{calc} - \delta, \phi_\text{calc} + \delta]$ and look at the purity of the final state. Here, $\delta$ is a phase interval which determines the trade-off between purity of the chosen states and the number of successful kick backs.

%Numerical simulations suggest that the effect of the feedback phase uncertainty $\delta$ is a less than 1\% decrease of $\bar{P}$ for  $\delta < \unit[10]{degrees}$. $\varepsilon_d$ should set a lower limit for a clear effect. WHAT IS MEANT BY THIS LAST SENTENCE?

%\subsection{Other parameters}

%In \figref{PurityChi} we plot how varying $\chi$ affects how well the dephasing can be undone. We see that $\chi < 2\kappa$ seems to be the optimal value for observing a clear difference between doing phase correction and not doing it. The oscillations in the figure are due to the measurement time being the same for all points; the oscillations in \figref{PurityTime} depend on $\chi$. We also note that a large $\chi$ diminishes the need for undoing dephasing, which was shown in Ref.~\cite{LalumierePRA2010}.

%\begin{figure}\centering
%\includegraphics[width=\linewidth]{Figures/VaryChiPaper120125.eps}
%\caption[]{The purity of the qubit density matrix after measurement, with no phase correction (red dots), with phase correction after $t=t_\text{meas}$ (blue dots), and with phase correction after $t=t_\text{meas} +t_\text{off} $ (green dots). The same parameters as in \figref{PurityTime}, except for $t_\text{off} = 10$ and $\chi$, which is varied.} \label{PurityChi}
%\end{figure}

%Numerical simulations suggest that the effect of the feedback phase uncertainty $\delta$ is a less than 1\% decrease of $\bar{P}$ for  $\delta < \unit[10]{degrees}$. $\varepsilon_d$ should set a lower limit for a clear effect.

%%%%%%%%%%%%%%%%%%%%%%%%%%%%%%%%%%%

\section{Conclusion}\label{sec_conclusion}

We have studied the measurement-induced dephasing in a single qubit in the two cases of homodyne detection and photodetection. We show that the dephasing can be completely undone if all the photons in the resonator are measured. We also show that this procedure undoes all the unwanted backaction in a parity measurement on two qubits, thereby improving the results of Ref.~\cite{TornbergPRA2010}. Furthermore, we have described and analyzed an experimental proposal for testing the effect in the case of homodyne detection on one qubit in a resonator. We show that the experiment is feasible in a parameter regime accessible with existing technologies in circuit QED.

%%%%%%%%%%%%%%%%%%%%%%%%%%%%%%%%%%%

\section{Acknowledgments}\label{sec_ack}
We thank L. DiCarlo, H. Wiseman, A. Korotkov, and H. Mabuchi for valuable discussions. This work is supported by the Swedish Research Council, and by the EU through the integrated project SOLID.

%%%%%%%%%%%%%%%%%%%%%%%%%%%%%%%%%%%

\appendix*
\section{Deriving the effective stochastic master equation for photodetection}\label{app_A}
To derive the effective SME for photodetection, \eqref{EffectiveSMEPhoto}, from the full SME, \eqref{UntransfSMEphoto}, we follow the procedure outlined in Ref.~\cite{GambettaPRA2008} which was used to derive \eqref{EffectiveSMEHom} from \eqref{UntransfSMEHom}. We begin by applying the displacement transformation, \eqref{DisplacementTransf}, on the unnormalized version of \eqref{UntransfSMEphoto} (replacing the superoperators $\mathcal{M}$ and $\mathcal{G}$ with $\measb{c}\rhob = c\rhob + \rhob c^\dag$ and $\measgb{c}\rhob = c\rhob c^\dag - \rhob$)
\begin{eqnarray}
\rd \rhobp &=& -\frac{i}{\hbar}\comm{H_\text{eff}^\bP}{\rhobp}\rd t + (1-\eta)\kappa\lind{a^\bP}\rhobp \rd t \nonumber\\
&&+ \gamma_1\lind{\sigma_-^\bP}\rhobp \rd t + \frac{\gamma_\phi}{2} \lind{\sigma_z^\bP}\rhobp \rd t \nonumber\\
&&- \frac{1}{2}\eta\kappa \measb{a^{\dag\bP} a^\bP}\rhobp \rd t + \measgb{a^\bP}\rhobp \rd N(t) \nonumber\\
&& +  \rhobp\bP^\dag\rd\bP + \rd\bP^\dag \bP\rhobp, 
\label{RhoPDot}
\end{eqnarray}
where we have defined operators in the displaced frame $c^\bP = \bP^\dag c \bP$. By a direct calculation of the transformed operators
%%
%\begin{eqnarray}
%\dot{\rho}^\bP &=& -i\comm{\frac{\tilde{\omega}_a}{2}\sigma_z + \left(\Delta_r + \chi\sigma_z \right)a^\dag a}{\rhop} + (1-\eta)\kappa\lind{a}\rhop \nonumber \\
%&&+ \gamma_1\lind{\sigma_-^\bP}\rhop + \frac{\gamma_\phi}{2} \lind{\sigma_z}\rhop \nonumber \\
%&&+ (1-\eta)\kappa\left(\lind{\Pi_\alpha}\rhop + a\comm{\rhop}{\Pi_\alpha^\dag} + \comm{\Pi_\alpha}{\rhop}a^\dag\right) \nonumber\\
%&&+ \Bigg[- i\left(\varepsilon_d (t)\left(a^\dag + \Pi_\alpha^\dag\right) + \varepsilon_d^* (t) \left(a + \Pi_\alpha \right)\right) \nonumber\\
%&&\quad\:\:\: - i\left(\Delta_r + \chi\sigma_z \right)\left(a^\dag\Pi_\alpha + a\Pi_\alpha^\dag + \Pi_\alpha^\dag\Pi_\alpha\right)  \nonumber\\
%&&\quad\:\:\:  - \dot{\Pi}_\alpha a^\dag +  \dot{\Pi}_\alpha^\dag a - i\sum_{i=e,g}\Pi_i\im\left(\dot{\alpha_i}\alpha_i^*\right) \nonumber\\
%&&\quad\:\:\:  + (1-\eta)\frac{\kappa}{2} \left(a\Pi_\alpha^\dag - a^\dag\Pi_\alpha \right), \rhop\Bigg]. \label{monster}
%\end{eqnarray}
%%
and using Eqs.~(\ref{AlphaGDot}) and (\ref{AlphaEDot}) we arrive at the SME in the displaced frame
\begin{widetext}
\begin{eqnarray}
\rd \rhobp &=& -i\comm{\frac{\tilde{\omega}_a}{2}\sigma_z + \left(\Delta_r + \chi\sigma_z \right)a^\dag a + \frac{1}{2}\left( \varepsilon_d \Pi_\alpha^\dag + \varepsilon_d^* \Pi_\alpha \right)}{\rhobp}\rd t + \left(1-\eta\right)\kappa\lind{a}\rhobp \rd t + \gamma_1\lind{\sigma_-^\bP}\rhobp \rd t  \nonumber \\
&&+ \frac{\gamma_\phi}{2} \lind{\sigma_z}\rhobp \rd t + \left(1-\eta\right)\kappa \left(\lind{\Pi_\alpha}\rhobp + a\comm{\rhobp}{\Pi_\alpha^\dag} + \comm{\Pi_\alpha}{\rhobp}a^\dag\right) \rd t \nonumber\\
&&- \frac{1}{2}\eta\kappa\left(\left(a^\dag a + 2a\Piad + \Piad\Pia\right)\rhobp + \rhobp\left(a^\dag a + 2a^\dag \Pia + \Piad\Pia\right) \right)\rd t \nonumber\\
&&+ \bigg(a\rhobp a^\dag + a\rhobp\Piad + \Pia\rhobp a^\dag + \Pia\rhobp\Piad - \rhobp \bigg)\rd N(t).
\label{DRhoPTransformed}
\end{eqnarray}
\end{widetext}

The transformed density matrix can be expanded in eigenstates of $\sigma_z$ and $a^\dag a$
\begin{equation}
\rhobp = \sum_{n,m = 0}^\infty \sum_{i,j = g,e} \rhobp_{nmij}\ket{n,i}\bra{m,j},
\end{equation}
such that the reduced density matrix for the qubit, $\rhoq$, is found by tracing out the resonator degrees of freedom
\begin{eqnarray}
\rhobq &=& \text{tr}_{\text{res}}\left(\bP\rhobp\bP^\dag\right) \\
&=& \sum_{n,m = 0}^\infty \sum_{i,j = g,e}\rhobp_{nmij}\ket{i}\bra{j}\left\bra{m}D^\dag\left(\alpha_j\right) D\left(\alpha_i\right)\right\ket{n} \nonumber\\ 
&=& \sum_{n,m = 0}^\infty \sum_{i,j = g,e}\rhobp_{nmij}\ket{i}\bra{j}d_{mnij}\exp\left(-i\im\left(\alpha_j\alpha_i^*\right)\right),\nonumber
\end{eqnarray}
where we have defined $d_{pqij} = \brakket{p}{D\left(\beta_{ij}\right)}{q} $, and $\beta_{ij} = \alpha_i - \alpha_j$.

The evolution of the density matrix elements are given by \eqref{DRhoPTransformed}
\begin{widetext}
\begin{eqnarray}
\rd\rhobp_{nmgg} &=& \bigg(\left(-i\left(\Delta_r - \chi\right)(n-m) - \frac{\kappa}{2}(n+m)\right)\rhobp_{nmgg} + \kappa\sqrt{(n+1)(m+1)}\rhobp_{n+1,m+1,gg}  + \gamma_1\sum_{p,q = 0}^\infty \rhobp_{pqee}d_{npeg}d_{qmge} \nonumber\\
&&- \frac{1}{2}\kappa\eta\left(\left(n + m + 2\abs{\alpha_g}^2 \right)\rhobp_{nmgg} + \alpha_g^*\sqrt{n+1}\rhobp_{n+1,mgg} + \alpha_g\sqrt{m+1}\rhobp_{n,m+1,gg} \right)\bigg)\rd t \\
&&+ \left(\sqrt{(n+1)(m+1)}\rhobp_{n+1,m+1,gg} + \alpha_g^*\sqrt{n+1}\rhobp_{n+1,mgg} + \alpha_g\sqrt{m+1}\rhobp_{n,m+1,gg} +  \left(\abs{\alpha_g}^2 - 1\right)\rhobp_{nmgg}\right)\rd N(t),\nonumber
\label{rhop_nmgg_dot}
\end{eqnarray}
\begin{eqnarray}
\rd\rhobp_{nmee} &=& \bigg(\left(-i\left(\Delta_r - \chi\right)(n-m) - \frac{\kappa}{2}(n+m)\right)\rhobp_{nmee} + \kappa\sqrt{(n+1)(m+1)}\rhobp_{n+1,m+1,ee} + \gamma_1\rhobp_{nmee} \nonumber\\
&&- \frac{1}{2}\kappa\eta\left(\left(n + m + 2\abs{\alpha_e}^2 \right)\rhobp_{nmee} + \alpha_e^*\sqrt{n+1}\rhobp_{n+1,mee} + \alpha_e\sqrt{m+1}\rhobp_{n,m+1,ee} \right)\bigg)\rd t\\
&&+ \left(\sqrt{(n+1)(m+1)}\rhobp_{n+1,m+1,ee} + \alpha_e^*\sqrt{n+1}\rhobp_{n+1,mee} + \alpha_e\sqrt{m+1}\rhobp_{n,m+1,ee}  +  \left(\abs{\alpha_e}^2 - 1\right)\rhobp_{nmee}\right)\rd N(t),\nonumber
\label{rhop_nmee_dot}
\end{eqnarray}
\begin{eqnarray}
\rd \rhobp_{nmeg} &=& \Bigg(-i\left(\tilde{\omega}_a + (n-m)\Delta_r + (n+m)\chi + \frac{1}{2}\left(\varepsilon_d\beta_{eg}^* + \varepsilon_d^*\beta_{eg} \right)\right) - \frac{(1-\eta)\kappa}{2}(n+m)  \nonumber\\
& &- \frac{\gamma_1}{2} - \gamma_\phi - (1-\eta)\kappa\Bigg(\frac{\abs{\beta_{eg}}^2}{2} + i\im(\alpha_e^*\alpha_g)\Bigg) \Bigg)\rhobp_{nmeg}\rd t \nonumber\\ 
&&+ (1-\eta)\kappa\left(\sqrt{(n+1)(m+1)}\rhobp_{n+1,m+1,eg} + \beta_{eg}\sqrt{m+1}\rhobp_{n,m+1,eg} - \beta_{eg}^*\sqrt{n+1}\rhobp_{n+1,meg} \right)\rd t \\ 
&&- \frac{\eta\kappa}{2}\bigg( \left(n + m + \abs{\alpha_g}^2 + \abs{\alpha_e}^2 \right)\rhobp_{nmeg} +\alpha_e^*\sqrt{n+1}\rhobp_{n+1,meg} + \alpha_g\sqrt{m+1}\rhobp_{n,m+1,eg} \bigg)\rd t \nonumber\\
&&+ \left(\sqrt{(n+1)(m+1)}\rhobp_{n+1,m+1,eg} + \alpha_g^*\sqrt{n+1}\rhobp_{n+1,meg} + \alpha_e\sqrt{m+1}\rhobp_{n,m+1,eg} + \left(\alpha_e\alpha_g^* - 1\right) \rhobp_{nmeg}\right)\rd N(t).\nonumber
\label{rhop_nmeg_dotPhoto}
\end{eqnarray}
\end{widetext}

Noting that there are no mechanisms to populate higher photon states in Eqs.~(\ref{rhop_nmgg_dot}) and (\ref{rhop_nmee_dot}), and that we have transformed to a frame where the initial photon population is zero, we arrive at the equations for the diagonal elements of the qubit density matrix:
\begin{eqnarray}\label{rhoq_ii_DotPhoto}
\rd \rhobq_{ii} = \rd \rhobp_{00ii} &=& \pm\gamma_1\rhobq_{ee}\rd t - \kappa\eta\abs{\alpha_i}^2\rhobq_{ii}\rd t \nonumber\\
&&+ \left(\abs{\alpha_i}^2 - 1\right)\rhobq_{ii}\rd N(t),
\end{eqnarray}
where the + (-) in the first term should be used when $i=g$ ($i=e$). For the off-diagonal elements we have 
\begin{eqnarray}
\rd \rhobq_{eg} = \sum_{n,m} \rd \bar{\lambda}_{nmmneg},
\end{eqnarray}
where
\begin{equation}
\bar{\lambda}_{nmpqij} = \rhobp_{nmij}d_{pqij}\exp\left(-i\im\left(\alpha_j\alpha_i^*\right)\right).
\end{equation}
To simplify this expression we look at the evolution of the $\bar{\lambda}$'s. Using that
\begin{eqnarray}
\dot{d}_{pqeg} &=& \left(\frac{\kappa}{2}\left| \beta_{eg}\right|^2 - 2\chi\im\left(\alpha_g\alpha_e^*\right)\right)d_{pqeg}\nonumber\\
&&+ \dot{\beta}_{eg}\sqrt{p} d_{p-1,qeg} \nonumber\\
&&- \dot{\beta}_{eg}^*\sqrt{q} d_{p,q-1,eg}  ,  \label{ddot} \\
\frac{\partial}{\partial t}\left(\im\left(\alpha_g\alpha_e^*\right)\right) &=& -\frac{1}{2}\left(\varepsilon_d\beta_{eg}^* + \varepsilon_d^*\beta_{eg} \right) + 2\chi\re\left(\alpha_g\alpha_e^*\right) \nonumber\\
&&- \kappa\im\left(\alpha_e^*\alpha_g\right) \label{expdot},
\end{eqnarray}
and inserting the result from \eqref{rhop_nmeg_dotPhoto} gives us 
\begin{widetext}
\begin{eqnarray}
\rd \bar{\lambda}_{nmpqeg} &=& \bigg(-i\left(\tilde{\omega}_a + B - \eta\kappa\im\left(\alpha_e^*\alpha_g\right) \right) - i(n-m)\Delta_r - i(n+m)\chi \nonumber\\
&& - \frac{(1-\eta)\kappa}{2}(n+m) - \frac{\gamma_1}{2} - \gamma_\phi - \Gamma_d + \frac{\eta\kappa}{2}\left| \beta_{eg}\right|^2 \bigg)\bar{\lambda}_{nmpqeg}\rd t \nonumber\\
&& + (1-\eta)\kappa\sqrt{(n+1)(m+1)}\bar{\lambda}_{n+1,m+1,pqeg}\rd t + (1-\eta)\kappa\beta_{eg}\sqrt{m+1}\bar{\lambda}_{n,m+1,pqeg}\rd t \nonumber\\
&& - (1-\eta)\kappa\beta_{eg}^*\sqrt{n+1}\bar{\lambda}_{n+1,mpqeg}\rd t  +  \dot{\beta}_{eg}\sqrt{p}\bar{\lambda}_{nm,p-1,qeg}\rd t - \dot{\beta}_{eg}^*\sqrt{q}\bar{\lambda}_{nmp,q-1,eg}\rd t  \nonumber\\
&& - \frac{\eta\kappa}{2}\left( \left(n + m + \abs{\alpha_g}^2 + \abs{\alpha_e}^2 \right)\bar{\lambda}_{nmpqeg} +\alpha_e^*\sqrt{n+1}\bar{\lambda}_{n+1,mpqeg} + \alpha_g\sqrt{m+1}\bar{\lambda}_{n,m+1,pqeg} \right)\rd t \nonumber\\
&& + \bigg(\sqrt{(n+1)(m+1)}\bar{\lambda}_{n+1,m+1,pqeg} + \alpha_g^*\sqrt{n+1}\bar{\lambda}_{n+1,mpqeg} \nonumber\\
&& + \alpha_e\sqrt{m+1}\bar{\lambda}_{n,m+1,pqeg} +  \left(\alpha_e\alpha_g^* - 1\right)\bar{\lambda}_{nmpqeg}\bigg)\rd N(t).
\label{lambdadot2Photo}
\end{eqnarray}
\end{widetext}

Noting again that there are no mechanisms to populate higher photon states, and that the initial photon population is zero in the transformed frame, we get an equation for the off-diagonal element of the reduced density matrix:
\begin{eqnarray}
\rd \rhobq_{eg} = \rd \bar{\lambda}_{0000eg} &=& \Big(-i\left(\tilde{\omega}_a + B - \eta\kappa\im\left(\alpha_e^*\alpha_g\right)\right) - \frac{\gamma_1}{2} \nonumber\\
&& - \gamma_\phi - \Gamma_d + \frac{\eta\kappa}{2}\left| \beta_{eg}\right|^2\Big) \rhobq_{eg} \rd t\nonumber\\
&& - \frac{\eta\kappa}{2}\left(\abs{\alpha_g}^2 + \abs{\alpha_e}^2 \right)\rhobq_{eg} \rd t \nonumber\\
&& + \left(\alpha_e\alpha_g^* - 1\right)\rhobq_{eg}\rd N(t).
\label{rhoqegDotPhoto}
\end{eqnarray}

From Eqs.~(\ref{rhoq_ii_DotPhoto}), (\ref{rhoqegDotPhoto}) and the identity
\begin{eqnarray}
&&\frac{i\eta\kappa}{2}\im\left(\alpha_e^*\alpha_g\right)\comm{\sigma_z}{\rhoq} - \frac{\eta\kappa}{4}\left| \beta_{eg}\right|^2\lind{\sigma_z}\rhoq \nonumber\\
&=& -\eta\kappa\lind{\Pi_\alpha}\rhoq,
\label{DPiaIdentity}
\end{eqnarray}
we get the effective SME for photodetection
\begin{eqnarray}
\rd\rhobq &=& -i\frac{\tilde{\omega}_a + B}{2}\comm{\sigma_z}{\rhobq}\rd t + \gamma_1\lind{\sigma_-}\rhobq\rd t \\
&&+ \frac{\gamma_\phi + \Gamma_d}{2}\lind{\sigma_z}\rhobq \rd t - \eta\kappa\lind{\Pi_\alpha}\rhobq\rd t \nonumber\\
&& - \frac{1}{2}\eta\kappa\measb{\Pi_\alpha^\dag\Pi_\alpha}\rhobq\rd t + \measgb{\Pi_\alpha}\rhobq\rd N(t),\nonumber
\label{EffectiveSMEPhotoAppendix}
\end{eqnarray}
where $\expecc{\rd N(t)} = \eta\kappa\expec{\Pia^\dag\Pia}\rd t$. This is exactly what one gets by simply replacing $a$ with $\Pia$ in all the measurement terms of the original SME. 

Normalizing \eqref{EffectiveSMEPhotoAppendix} gives \eqref{EffectiveSMEPhoto}. Note that in the limit $\eta \rightarrow 0$ \eqref{EffectiveSMEPhoto} reduces to the effective ME without measurement. Secondly, averaging over an ensemble of quantum trajectories gives
\begin{eqnarray}
&&\expecc{-\frac{1}{2}\eta\kappa\meas{\Pi_\alpha^\dag\Pi_\alpha}\rhoq\rd t + \measg{\Pi_\alpha}\rhoq\rd N(t)} \nonumber\\
&=& \eta\kappa\lind{\Pi_\alpha}\rhoq\rd t,
\end{eqnarray}
which lets us recover the effective ME without measurement.

%%%%%%%%%%%%%%%%%%%%%%%%%%%%%%%

\end{document}